# Temperature Dependence Calibration and Correction of the DAMPE BGO Electromagnetic Calorimeter

Yifeng Wei, Zhiyong Zhang, Yunlong Zhang*, Sicheng Wen, Chi Wang, Zhiying Li, Changqing Feng, Xiaolian Wang, Zizong Xu, Guangshun Huang*, Shubin Liu

*State Key Laboratory of Particle Detection and Electronics, University of Science and Technology of China, Hefei, Anhui, China, 230026*

*Department of Modern Physics, University of Science and Technology of China, Hefei, Anhui, China, 230026*

*Abstract–* **A BGO electromagnetic calorimeter is built for DArk Matter Particle Explorer(DAMPE) mission. The temperature effect of BGO ECAL has been investigated with a thermal vacuum experiment. The light output of BGO crystal depends on temperature significantly. Temperature coefficient of each BGO crystal bar has been calibrated, and a correction method is also presented in this paper.**

## 1. Introduction

The DArk Matter Particle Explorer project is a Chinese space detection mission for precise measurement of high energy electron, gamma ray and nuclei from deep space. It has been launched in the end of 2015. So far, the physical origin of unexpected galactic electron/positron excesses [1][2][3] is still to be figured out. DAMPE will provide a high energy resolution measurement of ($e^+ + e^-$) spectrum from 5 GeV to 10 TeV, which has not been revealed by existing experiments in 1 TeV to 10 TeV ranges. DAMPE can also measure protons, helium nuclei and other cosmic ions in the energy band from 10 GeV to ~ 1 PeV [4].

The DAMPE detector consists of four sub-detectors (Fig. 1). The plastic scintillator detector(PSD) is used to identify electron, photon and cosmic ions; the silicon tungsten tracker(STK) provides tracking; and the neutron detector(NUD) is designed to improve the electron/proton separating capability for proton background rejection in space. The emphasis of this paper is the BGO electromagnetic calorimeter (BGO ECAL), which provides a high precision measurement of energy.

The DAMPE satellite orbits the earth at the altitude of 500 km, and the designed maximum temperature variation of the BGO ECAL ranges from -15$^o$C to +20$^o$C under the satellite thermal control system. The ECAL is made up of 308 BGO crystal bars, each with dimensions of 25×25×600 mm$^3$. Like most of inorganic crystals, the scintillation light yield of BGO crystal depends on the temperature. The reference value of temperature dependence of BGO crystal is -0.9%/$^o$C [5], and the PMTs produced by Hamamatsu also depend on temperature (~ -0.4%/$^o$C) [6]. We have investigated the temperature dependence of a single BGO crystal bar coupled with a PMT [7]. In order to well understand the temperature performance of the full calorimeter, a thermal vacuum experiment of DAMPE flight model has been performed in July, 2015.

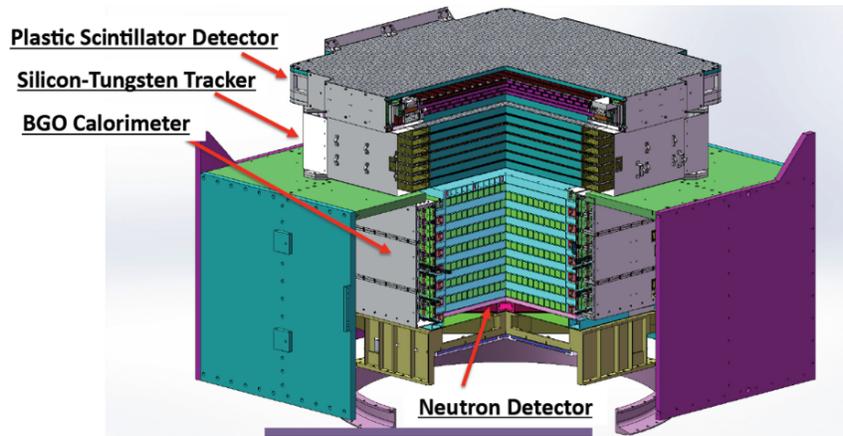

Fig. 1 Structure of DAMPE detector

## 2. Design of BGO Calorimeter

The BGO ECAL is to measure energy deposition of particles with a wide range, from 5 GeV to 10 TeV, and with a good energy resolution of 1.5% at 800 GeV for photons and electrons. The ECAL contains 14 layers (~ 31 radiation lengths), and 22 BGO crystal bars in each layer. The layers of BGO crystal bars are alternated in an orthogonal way to measure the deposited energy and profile of hadron and electromagnetic showers developed in the BGO ECAL (Fig. 2).

Scintillation light is detected at both ends of a BGO crystal bar with two PMTs, respectively. An asymmetry design is applied in the readout of two PMTs. The signal of high gain output (called Side 0) is about five times as large as the low gain (Side 1). The required range of energy response for one BGO detector unit (BGO crystal bar + PMT) is from 10 MeV to 2 TeV. To obtain such a wide dynamic range, a multi-dynode readout circuit of PMT base is used, and dynode 2, 5, 8 of PMT correspond to low, middle and high gain, respectively [8][9].

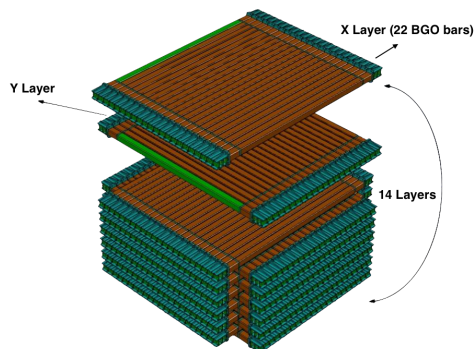

Fig. 2  BGO crystal bars arrangement in the ECAL

## 3. Thermal vacuum experiment

The primary goal of thermal vacuum (TV) environmental experiment was to validate the thermal performance of the DAMPE flight model. For the BGO sub-detector, the cosmic ray data in the period of TV experiment was utilized to study temperature dependence. The DAMPE detector was held in a cylinder chamber, and covered by a heating cage (Fig. 3). A liquid nitrogen cooling layer was around the instrument inside the chamber. There was a thermally insulating layer between the DAMPE detector and the support platform.

Fig. 4 shows the temperature variation measured by a thermistor attached on the surface of a BGO crystal bar, and the precision of this kind of thermistor is 0.1°C. There are one long cycle (July $2^{nd}$ ~ July $10^{th}$) and four short cycles (July $10^{th}$ ~ July $16^{th}$) during the TV experiment. The change rate of temperature in the large cycle, which was no more than 1°C/hour, was much gentler than the rate in the short cycle, and therefore, temperature dependence was calibrated with the data of the long cycle. Since two high voltage working modes (800V and 1000V) are applied on orbit (800V is normal mode), the experiment was carried on these two modes, respectively. The periods of different high voltage modes are marked in Fig. 4.

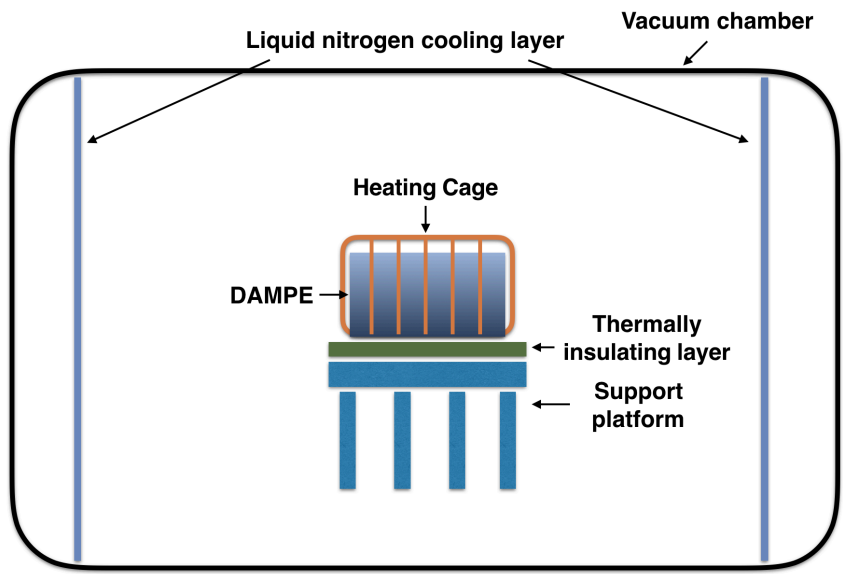

Fig. 3 Thermal vacuum experiment setup

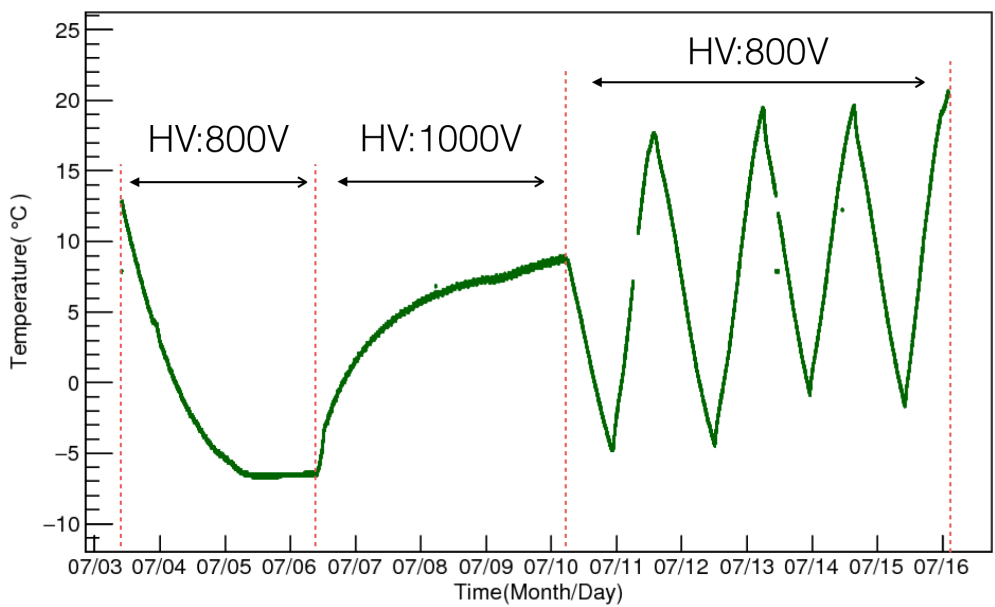

Fig. 4 Temperature curve of one BGO crystal bar

## 4. Temperature dependence calibration

The cosmic ray muon in the long cycle was used for temperature calibration. The energy deposit from a muon minimum ionizing particle (MIP) passing through a 2.5 cm thick BGO bar is about 22.5 MeV. Fig. 5 shows the energy deposition spectra of muon MIPs at different temperatures. These spectra were fitted with Landau convoluted with Gaussian, and the Landau MPV values could represent the light output of BGO crystal bars. For an individual detector unit (BGO crystal bar + PMT), the MPV value decreases with the increase of temperature (Fig. 6), whose variation trend can be described well by a linear function in this temperature range. The temperature coefficient is defined as:

$$C_t = \frac{Slope\ of\ fit\ line}{Intercept\ of\ fit\ line} \quad (1)$$

This coefficient is exactly the temperature dependence at 0°C. Fig. 7 shows the distribution of $C_t$ values from the 308 BGO detector units, with a mean value of -1.322%/°C, which includes the contribution of BGO crystal as well as that of readout electronics (PMT, base board and FEE). In comparison, the temperature coefficient measured by ATIC experiment is -1.86%/°C [10].

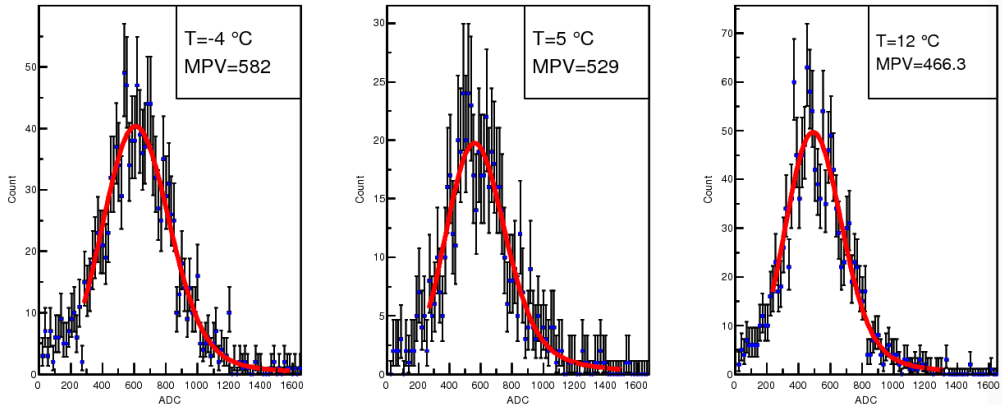

Fig. 5 Cosmic ray muon spectra at different temperatures

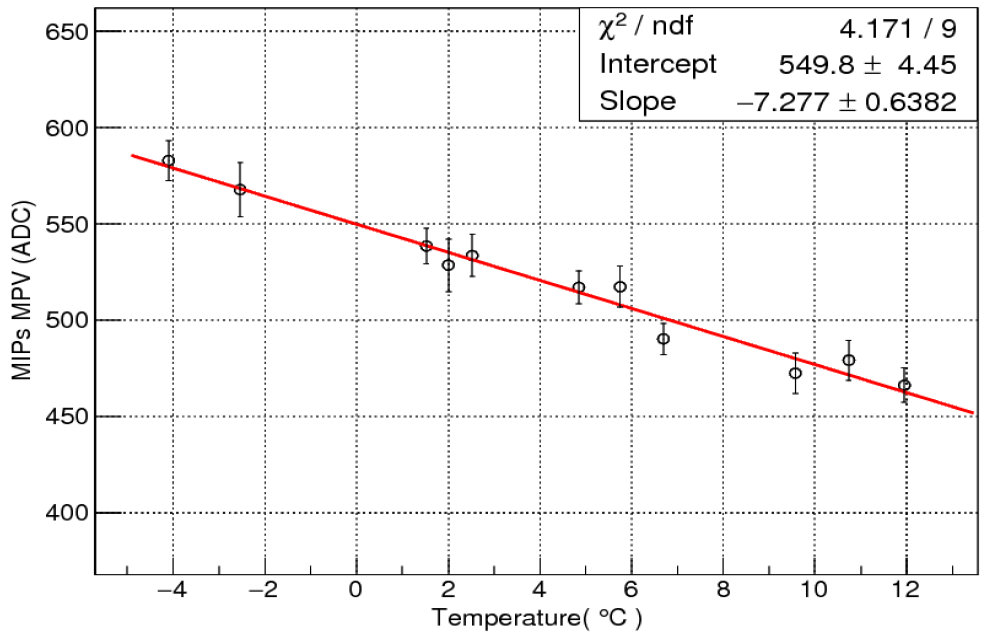

Fig. 6 The variation of MIPs MPV values with temperature

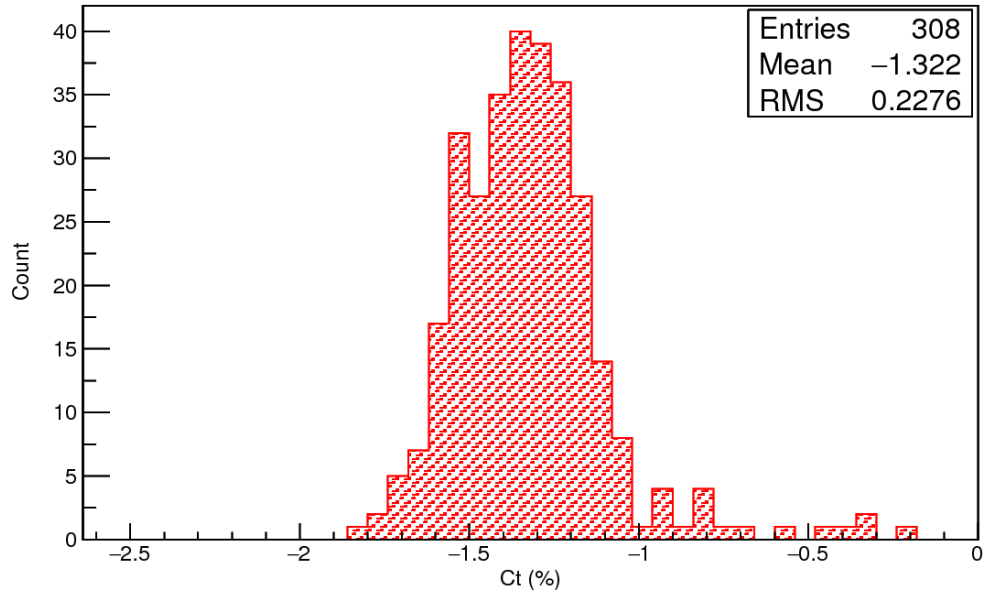

Fig. 7 Temperature coefficient distribution of 308 BGO crystal unit

## 5. Temperature effect correction

After calibrating the temperature coefficient of each BGO crystal unit, another point should be made clear. Only dynode 8 (high gain) can produce a significant signal when a muon MIP deposits its energy in a BGO crystal bar, because the ADC counts of dynode 2 and 5 are too small to be separated from pedestal noise, and this means, the coefficient calibrated with muon MIPs can only represent the temperature dependence of dynode 8. Therefore, the temperature dependence of dynode ratio of dy8 ADC to dy5 ADC has been investigated, which was calibrated by rare shower events of cosmic ray muon. Fig. 8 gives the change of dynode 8/dynode 5 ratio with time. It is clear that the dynode 8/dynode 5 ratio has no obvious temperature dependence, and the plateau is just because the high voltage was switched to 1000V mode during July.7 to July.10. Thus, the temperature coefficient calibrated with muon MIPs for dy8 could be used for other dynodes of the same PMT.

In correction algorithm, the following function is applied:

$$ADC_{Cor} = \frac{ADC_{raw}}{C_t \times T + 1} \quad (2)$$

Where $ADC_{raw}$ is raw ADC output of front electronics, $ADC_{Cor}$ is the value after temperature correction, T is the temperature and $C_t$ is the temperature coefficient.

This correction was applied to each output of the three PMT dynodes to compensate for temperature variation event by event, and the ADC counts of all channels were normalized to 0$^o$C.

The cosmic ray in the period of four short thermal cycles (in Fig. 4) is utilized for validation of temperature correction. Fig. 9 shows that the raw MPV values (triangle points) increase synchronously with the temperature decrease and after correction, the MPV values (circle points) approach to a constant at reference temperature (at 0$^o$C).

To evaluate the correction, the uniformity of MPV value is defined as:

$$\text{Uniformity} = \frac{RMS_{MPV}}{Mean_{MPV}} \quad (3)$$

Where RMS and Mean are statistical parameters of MIPs MPV values measured in different temperature.

The MPV uniformities of 308 BGO crystal units are shown in Fig. 10, which are improved from about 9% to about 2% after correction.

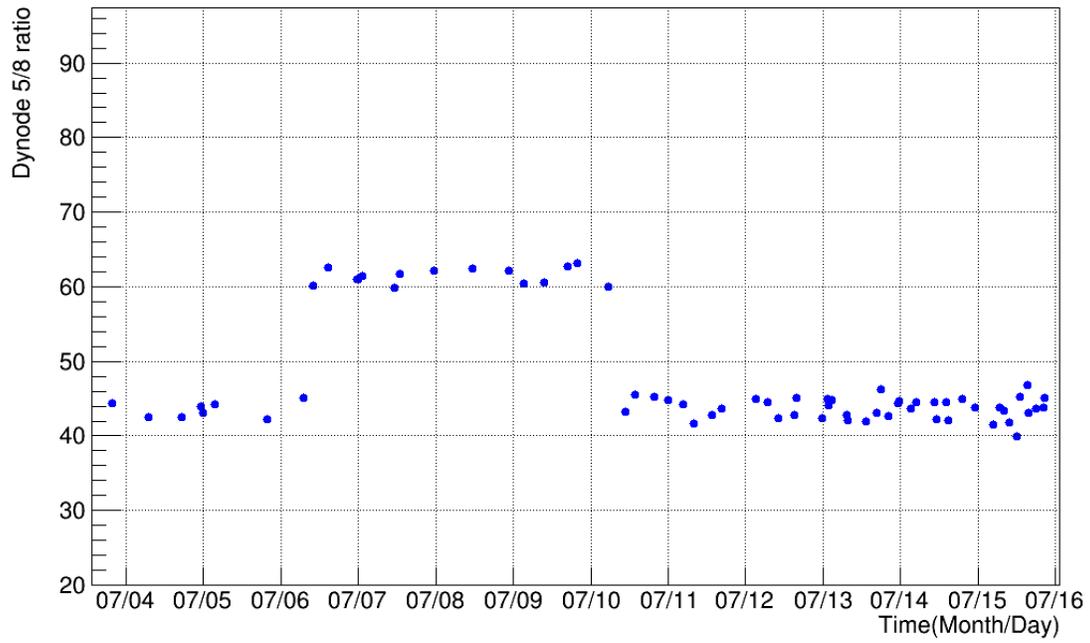

Fig. 8 Dynode 5/8 ratio with time

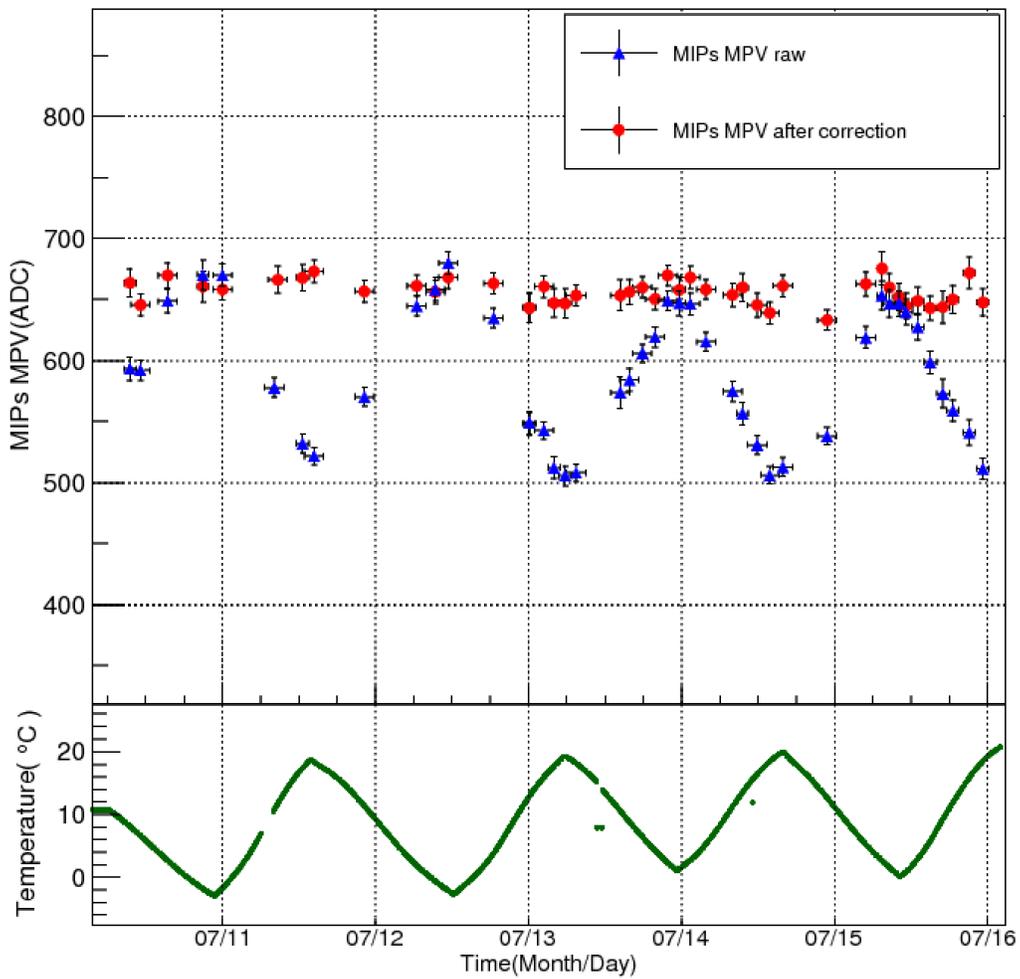

Fig. 9 Comparison of muon MIPs MPV value before and after correction

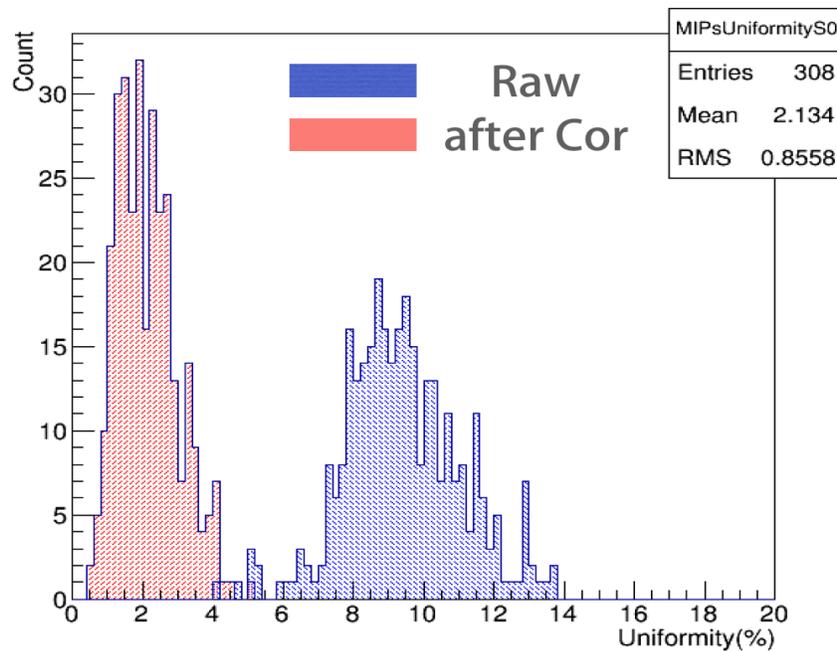

Fig. 10 Comparison of MPV uniformity before and after correction.

# 6. Conclusion

The temperature performance of DAMPE flight model BGO calorimeter has been studied in a thermal vacuum experiment. The mean value of temperature coefficients for the 308 units has been measured to be -1.322%. An algorithm has also been developed to correct the temperature effect, which improves the MPV uniformity to about 2%.

# Acknowledgments

This work was supported by the Chinese 973 Program, Grant No. 2010CB833002, the Strategic Priority Research Program on Space Science of the Chinese Academy of Science, Grant No. XDA04040202-4 and 100 Talents Program of CAS.

The author would like to thank to the colleagues in DAMPE collaboration, NUAA (Nanjing University of Aeronautics and Astronautics) and SECM (Shanghai Engineering Center for Microsatellites) for their contributions to the thermal vacuum experiment.